\documentstyle[prd,aps,epsfig]{revtex}

\def\tk{\tilde\kappa}
\def\tf{\tilde{f}}

\begin{document}

\thispagestyle{plain}
\setcounter{page}{1}

\title{On the Angular Dependence of the Radiative Gluon Spectrum}

\author{R.~Baier}
\address{
Fakult\"{a}t f\"{u}r Physik, Universit\"{a}t Bielefeld,
D-33501 Bielefeld, Germany}

\author{Yu.L.~Dokshitzer\footnote{
on leave from PNPI, Gatchina 188350, 
St.~Petersburg, Russia}}
\address{LPTHE\footnote{Laboratoire associ\'e du Centre National de 
la Recherche Scientifique},
 Universit\'e Pierre et Marie Curie (Paris VI), 4, Place Jussieu, 
F-75252 Paris, France,\\ and \\
LPT$^{\dagger}$,
 Universit\'e Paris-Sud, B\^atiment 210, F-91405 Orsay, France}

\author{A.H.~Mueller\footnote{Supported in part by the U.S. Department of 
Energy under Grant DE-FG02-94ER-40819}}
\address{Physics Department, Columbia University, New York, NY 10027, USA}

\author{D.~Schiff}\address{LPT$^{\dagger}$ , Universit\'e Paris-Sud,
 B\^atiment 210, F-91405 Orsay, France}

\maketitle

\begin{abstract}
The induced momentum spectrum of soft gluons radiated from a high energy 
quark produced in and  propagating through a QCD medium
 is reexamined in the BDMPS formalism.
A mistake in our published work (Physical Review C60 (1999) 064902)
is corrected.
The correct dependence of the fractional induced loss $R(\theta_{{\rm cone}})$
as a universal function  of the variable 
$\theta^2_{{\rm cone}} L^3 \hat q$ where $L$ is the size of the medium and 
$\hat q$ the transport coefficient is presented.
We add the proof that the radiated gluon momentum spectrum
derived in  our formalism  is  equivalent
with the one derived in the Zakharov-Wiedemann approach.
\end{abstract}


\section{Introduction}

In  the present paper we correct an error made in the calculation for
 the angular distribution of radiated
gluons as presented in \cite{Baier}. We recalculate the
quantitative predictions of
the integrated energy loss outside an angular cone, defining the jet,
with fixed opening angle, $\Delta E (\theta_{{\rm cone}})$,
and derive the expression for the ratio $R (\theta_{{\rm cone}} ) = 
{\Delta E (\theta_{{\rm cone}})/\Delta E}$, where 
$\Delta E$ is the completely integrated loss.
We confirm our previous
result  that $R (\theta_{{\rm cone}})$ is a universal function of the 
variable $\theta^2_{{\rm cone}} \hat q L^3$, where $\hat q$ is the 
transport coefficient characteristic of the medium. 
In particular we now find that  $R (\theta_{{\rm cone}} )$ is strongly
peaked at small values of  $\theta_{{\rm cone}}$
  ($\theta^2_{{\rm cone}} \hat q L^3 \simeq 1/10$),
a feature already obtained by Wiedemann \cite{Wiedemann}.
Finally, we explicitly show that the induced radiated gluon momentum spectrum
derived in  our formalism  is  equivalent
with the one derived in the Zakharov-Wiedemann approach
\cite{Zakharov}, namely 
in  the form given in\cite{Wiedemann}.

\section{Momentum spectrum of radiated gluons} \label{momentum spectrum}

In this section we repeat shortly the discussion of 
the induced momentum spectrum of soft gluon 
emission $(x\rightarrow 0)$ from a fast quark jet, which is 
produced {\it inside} the medium by a hard scattering at time $t=0$. 
From the production point the quark propagates over 
a length $L$ of QCD matter, 
carrying out many scatterings via gluon exchange with the medium. 

In the BDMPS approach\cite{BDMS} the spectrum per unit length $z$ of the 
medium consists of two terms: one corresponding to an 
``on-shell'' quark, as if that quark were entering the medium, and one 
corresponding to a hard production 
vertex.
From \cite{Baier} we quote
the induced gluon spectrum in the soft $\omega$ limit:
\begin{eqnarray}\label{eq:2.2}
\frac{\omega dI}{d\omega dz d^2 \underline{U}} & = & 
\frac{\alpha_s C_F}{\pi^2L} 2 \,Re \,\int^L_0 \, dt_2 \int d^2 \underline{Q}
 \left\{ 
\int^{t_2}_0 \, dt_1 \rho\sigma \frac{N_c}{2C_F} f (\underline{U} + 
\underline{Q} , t_2 - t_1 ) + f_h (\underline{U} + \underline{Q} , t_2) 
 \right\}  \nonumber \\
&\times& \rho\sigma \frac{N_c}{2 C_F} 2 \left[ \frac{\underline{U} + 
\underline{Q}}{(\underline{U} + \underline{Q})^2} - \frac{\underline{U}}
{\underline{U}^2}\right] V (\underline{Q} ) \,
{\cal F}_{fsi} \, 
\Bigg|^{\tilde\kappa =0}_
{\tilde\kappa},
\end{eqnarray}
where $f_h$ differs from $f$ in that the gluon emission time has been
evaluated at the time of the hard interaction, $t=0$.
We remind that the the medium-independent factorisation contribution
is eliminated by  a subtraction of the value of the integrals at 
$\tilde\kappa = 0$, where
this parameter is 
\begin{equation}\label{eq:2.3}
\tilde\kappa = \frac{2 C_F}{N_c} \frac{\lambda \mu^2}{2\omega}.
\end{equation}
It depends on medium properties, in particular
on the quark  mean free path $\lambda = 1/\rho\sigma$, as well as on the 
gluon energy $\omega$. The other symbols are explained in \cite{Baier}.

Next we  convert (\ref{eq:2.2}) to impact parameter space as explained in
\cite{Baier}, and we rescale the time variable
 $t = \frac{2 C_F}{N_c} \lambda\tau$ and
define
$\tau_0 = N_c L / 2 C_F \lambda$.
%
We obtain the 
spectrum in terms of impact parameter integrals, 
\begin{eqnarray}\label{eq:2.8}
\frac{\omega d I}{d\omega dz d^2 \underline{U}} & = & 
\frac{\alpha_s C_F}{\pi^2 L} 2 Re \, \int^{\tau_0}_0 \, d \tau_2 \, \int \,
\frac{d^2 B_1}{(2\pi)^2} \frac{d^2 B_2}{(2\pi)^2} \, 
e^{i(\underline{B}_1 - \underline{B}_2) \cdot \underline{U}} \,
\left\{ \int^{\tau_2}_0 \, d \tau_1 \tilde f (\underline{B}_1 , \tau_2 - 
\tau_1) +  \tilde f_h (\underline{B}_1 , 
\tau_2 ) \right\}  \nonumber \\
& \times & \frac{4 \pi i \underline{B}_2}{\underline{B}_2^2} \left[ \tilde V (
\underline{B}_1 - \underline{B}_2) - \tilde V ( \underline{B}_1) \right]
e^{- \frac{\tilde v}{2} (\underline{B}_1 - \underline{B}_2)^2 (\tau_0 - 
\tau_2)} \Bigg|^{\tilde\kappa = 0}_{\tilde\kappa} .
\end{eqnarray}
We still  have to evaluate the explicit 
expressions for  $\tilde f$, and especially for $\tilde f_h$.  
We recall that the amplitude $\tilde f (\underline{B} , \tau)$ is given 
by 
\begin{equation}\label{eq:2.9}
\tilde f (\underline{B}, \tau) = - \frac{i \pi \tilde v}{\cos^2 \omega_0 
\tau} \underline{B} \exp \left( - \frac{i}{2} m \omega_0 \underline{B}^2 
\tan \omega_0 \tau \right), 
\end{equation} 
with 
$m = - 1 / 2 \tilde\kappa \,\,\, {\rm and}\,\,\, \omega_0 = 
\sqrt{2i \tilde\kappa \tilde v}$. 
%
From the fact that $\tilde f_h$
and $\tilde f$ have the same time evolution
given by the harmonic oscillator Green function
$  G(\underline{B}_2,\tau_2;\underline{B}_1,\tau_1)$
 (cf. Eq.(5.6) of \cite{BDMS}),
i.e. in Eq. (\ref{eq:2.8}),
\begin{equation}\label{eq:2.xx}
  \tf_h(\underline{B}_1,\tau_2) \>=\> \int d^2B\>
 G(\underline{B}_1,\tau_2;\underline{B},\tau=0)\>
 \tf_h(\underline{B},\tau=0)\>,
\end{equation}
 while 
the initial conditions satisfy (in the soft gluon limit)
$\tilde f_h (
\underline{B} , 0) = \frac{2}{\tilde v \underline{B}^2} \tilde f 
(\underline{B},0) = -2\pi i {\underline{B}/ \underline{B}^2}$ as 
can be seen from Eqs.(17b) and (37)
of \cite{BDMS}.  

Here we note the mistake in our paper \cite{Baier}:
instead of the correct Eq.(\ref{eq:2.xx}) we have
taken 
\begin{equation}\label{eq:2.yy}
\tilde f_h (
\underline{B}_1 ,\tau)  = \frac{2}{\tilde v \underline{B}_1^2} \tilde f 
(\underline{B}_1,\tau) = \frac{2}{\tilde v \underline{B}_1^2}
\> \int d^2 B \>
 G(\underline{B}_1,\tau;\underline{B},\tau=0)\>
 \tf(\underline{B},\tau=0)\>,
\end{equation}
i.e. effectively we have taken out of the integrand in
Eq.(\ref{eq:2.yy}) the factor $1/{\underline{B}^2}$, and replaced it by
$1/{\underline{B}_1^2}$!

Eq.(\ref{eq:2.xx}) allows to derive the explicit time and impact 
parameter dependence of $\tilde f_h$,
which reads, 
\begin{equation}\label{eq:2.i}
\tilde f_h (\underline{B}, \tau) = - {2 \pi i} 
 {\underline{B}/ \underline{B}^2}\left[
 \exp \left( - \frac{i}{2} m \omega_0 \underline{B}^2 
\tan \omega_0 \tau \right) -
 \exp \left(\frac{i}{2} m \omega_0 \frac{\underline{B}^2} 
{\tan \omega_0 \tau} \right)  \right].
\end{equation} 
Based on this correct treatment of the hard production vertex 
the soft gluon radiation spectrum can be rewritten as
\begin{eqnarray}\label{eq:2.8i}
\frac{\omega d I}{d\omega dz d^2 \underline{U}} & = & 
- 4\frac{\alpha_s C_F \tilde{v}}{ L}  Re \, \int^{\tau_0}_0 \, d \tau \,
 \int \,
\frac{d^2 B_1}{(2\pi)^2} \frac{d^2 B_2}{(2\pi)^2} \, 
e^{i(\underline{B}_1 - \underline{B}_2) \cdot \underline{U}} \,
 \exp \left(\frac{i}{2} m \omega_0 \frac{\underline{B}_1^2} 
{\tan \omega_0 \tau} \right)  \nonumber \\
& \times & \frac{{\underline{B}_1} \cdot {\underline{B}_2}}
{{\underline{B}_1^2}{\underline{B}_2^2}}\,
 \left[ 
(\underline{B}_1  - \underline{B}_2)^2 -  \underline{B}_1^2 \right]\,
e^{- \frac{\tilde v}{2} (\underline{B}_1 - \underline{B}_2)^2 (\tau_0 - 
\tau)} \Bigg|^{\tilde\kappa = 0}_{\tilde\kappa} .
\end{eqnarray}
The $\underbar{B}$-space 
integral is performed as in \cite{Baier} 
and expressed  in terms of the function
\begin{equation}\label{eq:2.14}
J ( \underline{U} , \alpha , \beta ) = \frac{1}{16\pi^2}\, \frac{1}{\alpha 
(\alpha + \beta )} e^{-\frac{\underline{U}^2}{4 (\alpha + \beta)}}. 
\end{equation}

\section{Induced radiative energy loss of a hard quark jet in a finite cone}


In the following we recalculate the integrated loss 
{\it outside} an angular cone of opening angle $\theta_{{\rm cone}}$,
\begin{equation}\label{eq:3.1}
\Delta E (\theta_{{\rm cone}}) = L \, \int^\infty_0 \, d\omega\,
\int^\pi_{\theta_{{\rm cone}}} \, \frac{\omega dI}{d\omega dz d\theta} 
d\theta .
\end{equation}

\noindent
We note that for $\theta_{{\rm cone}} = 0$ the total loss 
is obtained, namely \cite{BDMS}
\begin{equation}\label{eq:3.2}
\Delta E = \frac{\alpha_s N_c}{4} \hat q L^2 . 
\end{equation}
We consider the normalized loss \cite{Baier} by defining the ratio 
\begin{equation}\label{eq:3.3}
R (\theta_{{\rm cone}} ) = \frac{\Delta E (\theta_{{\rm cone}})}
{\Delta E} ,
\end{equation}
with $R (\theta_{{\rm cone}}=0 ) = 1.$, by
using the same (dimensionless) variables and definitions as in \cite{Baier}.

We confirm that the ratio $R (\theta_{{\rm cone}})$
turns out to depend  on one single dimensionless variable
\begin{equation}\label{eq:3.8}
R = R (c (L) \theta_{{\rm cone}} ), 
\end{equation}
where
\begin{equation}\label{eq:3.9}
c^2 (L) = \frac{N_c}{2C_F} \hat q \left( L / 2 \right)^3 .
\end{equation}
This ``scaling behaviour'' of $R$
means that the medium and size dependence is universally contained in the
function $c (L)$, which is a function of the transport coefficient
 $\hat q = \tilde{v} \mu^2/\lambda$ 
 and of the length $L$, as defined by (\ref{eq:3.9}). 
 
As a consequence the discussion of the medium properties
is qualitatively the same as in \cite{Baier}, and does not 
need to be repeated here.
Quantitatively the corrected ratio $R$ is plotted in Fig.~1.

\begin{figure}
\centering
\epsfig{bbllx=35,bblly=210,bburx=515,bbury=625,
file=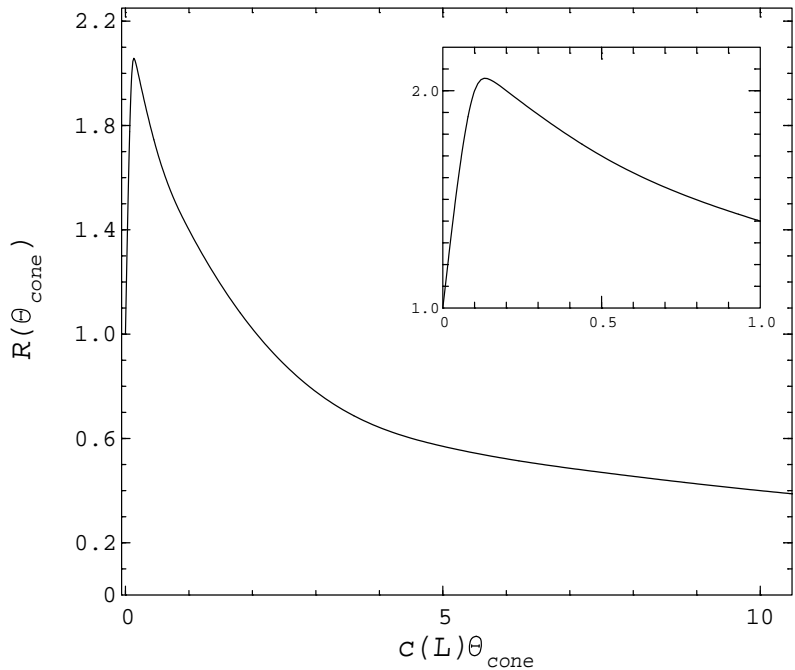,width=110mm,height=100mm}
\caption{\label{fig:ang}Fractional induced loss $R(\theta_{{\rm cone}})$
as a function of $ c(L)\theta_{{\rm cone}}$.}
\end{figure}

A sharp peak at small values of $ c(L)\theta_{{\rm cone}} \simeq 0.1 - 0.2$,
where  $R(\theta_{{\rm cone}}) \simeq 2.$,
can be seen in this figure, especially with the help of the insert. 
 This behaviour of the loss 
$\Delta E(\theta_{{\rm cone}})$
has been already noticed by Wiedemann in \cite{Wiedemann}.
For  $ c(L) \theta_{{\rm cone}} > 10$ the dependence is numerically 
as given in our previous work \cite{Baier}.

When comparing hot and cold QCD matter we recall  \cite{Baier} that
for fixed $L$
\begin{equation}\label{eq:3.13}
c (L) \Big|_{{\rm HOT}} \gg c(L) \big|_{{\rm COLD}}. 
\end{equation}
This implies that
e.g. for $L = 5 fm$, the position of the peak seen in Fig.~1
corresponds to very small cones:
 $\theta_{\rm cone}^{peak} \le 2^{\circ}$ for nuclear,
and $\theta_{\rm cone}^{peak} \le 0.5^{\circ}$ for hot
 (at reference temperature $T = 250 MeV$) matter, respectively.
For $\theta_{\rm cone} > \theta_{\rm cone}^{peak}$ the energy loss, 
 $R (\theta_{{\rm cone}})$ , drops quickly with increasing 
$\theta_{{\rm cone}}$, the behaviour already known from the
 discussion in \cite{Baier}.

For a quark jet produced in the medium  the  dependence
 of $R$ on $c(L) \theta_{{\rm cone}}$
is given for further reference,
 expressing explicitly the property Eq. (\ref{eq:3.8}) by

\begin{eqnarray}\label{eq:B.1}
R (\theta_{{\rm cone}}) &=& \frac{4}{\pi}\, Re \int^\infty_0\,
\frac{dx}{x^3} \, \int^1_0 \,y^2 dy \,\int^\infty_0\,dz \,
 \nonumber \\
& \times & \frac{1}{\left[ z - \frac{y}{(1+i)x\,\tan(1+i)x} \right]^2 } \,
\exp \left[ - \frac{c^2(L)\theta_{{\rm cone}}^2}{ 1 + z - y
 - \frac{y}{(1+i)x \,\tan(1+i)x} } \left( \frac{y}{x}\right)^4 \right] . 
\end{eqnarray}
The subtraction term in (\ref{eq:B.1})
  with $\tilde\kappa = 0$  vanishes. 

For the case of a quark produced outside the medium the calculation is the
same as in \cite{Baier}.

\section{Equivalence of the two formalisms}

We now show that Eq.~(\ref{eq:2.8}) can be expressed in a 
form identical to Eq.~(2.1), respectively Eqs.~(A.5) and (A.6)
of \cite{Wiedemann}. This derivation of the equivalence of our
formalism with the one by Wiedemann \cite{Wiedemann} and
Zakharov \cite{Zakharov} generalizes the proof given in \cite{BDMS},
but here for the full gluon momentum spectrum Eq.~(\ref{eq:2.2}).

As in \cite{BDMS} we express  
$\tilde f (\underline{B}_1, \tau_2 -\tau_1)$ and  
$\tilde f_h (\underline{B}_1, \tau_2)$ in terms of the Green function $G$,
as given by Eq.~(\ref{eq:2.xx}).
The function $G$ obeys the same differential equation as given for
$\tf$ \cite{BDMS},
\begin{equation} 
  \label{eq:44}
 \frac{\partial}{\partial \tau_1}
G(\underline{B}_1,\tau_2;\underline{B},\tau_1)\>=\> 
 i\tk\, \nabla^2_{\underline{B}} G 
+ \frac{\tilde v}{2}\underline{B}^2\, G\>.
\end{equation}
Using this Eq.~(\ref{eq:44}) we 
substitute the combination of $\partial G/\partial \tau_1$ and
$\nabla^2_{\underline{B}}G$ for this Green function
$ G$ which expresses the $\tau$ dependence of
  $\tilde f (\underline{B}_1, \tau_2 -\tau_1)$
in  Eq. (\ref{eq:2.8}). 
The lower limit of the $\tau_1$-integral of $\partial G/\partial \tau_1$ 
exactly cancels
the  term coming from $\tilde f_h (\underline{B}_1, \tau_2)$,
while its upper limit, at $\tau_2$, contains 
$$
G(\underline{B}_1,\tau_2;\underline{B},\tau_2) \>=\>
 \delta^2\left(\underline{B}_1-\underline{B} \right) ,
$$
which is  $\tk$-independent and therefore canceled by 
the $\tk\!=\!0$ subtraction term present in Eq.~(\ref{eq:2.8}).
Thus, we arrive at
\begin{eqnarray}
  \label{eq:45}
\frac{\omega d I}{d\omega dz d^2 \underline{U}} & = & 4
\frac{\alpha_s C_F}{L}\, 
 {\tilde\kappa \tilde v} Re (-i) \, \int^{\tau_0}_0 \, d \tau_2 \,
 \int^{\tau_2}_0 \, d \tau_1 \,
 \int \,\frac{d^2 B_1}{(2\pi)^2} \frac{d^2 B_2}{(2\pi)^2} \, 
 \int \,{d^2 B} \,
e^{i(\underline{B}_1 - \underline{B}_2) \cdot \underline{U}} \,
 \nonumber \\
& \times &
\frac{{\underline{B}} \cdot {\underline{B}_2}}
{{\underline{B}^2}{\underline{B}_2^2}}
  \left[ 
(\underline{B}_1  - \underline{B}_2)^2 -  \underline{B}_1^2 \right]
e^{- \frac{\tilde v}{2} (\underline{B}_1 - \underline{B}_2)^2 (\tau_0 - 
\tau_2)}
\, \nabla^2_{\underline{B}}
 G(\underline{B}_1,\tau_2;\underline{B},\tau_1) .
\end{eqnarray}
Using 
\begin{equation}
  \label{eq:50}
   \nabla_{\underline{B}} \cdot \, \frac{\underline{B}}{\underline{B}^2}
  \>=\> 2\pi \delta^2(\underline{B})\>,
\end{equation} 
and 
after integrating one $\nabla_{\underline{B}}$ derivative in Eq.~(\ref{eq:45})
 by parts one finds
\begin{equation}
  \label{eq:50a}
 \int \,{d^2 B} \,\frac{\underline{B}}{\underline{B}^2}\,
\nabla^2_{\underline{B}}\, G(\underline{B}_1,\tau_2;\underline{B},\tau_1)
 \>=\> - 2\pi \, \nabla_{\underline{B}} \,
 G(\underline{B}_1,\tau_2;\underline{B},\tau_1) . 
\end{equation} 
Now, expressing
\begin{equation}
  \label{eq:50b}
 \frac{2}{\tilde v}(\underline{B}_1  - \underline{B}_2)^2 =
 \frac{\partial}{\partial \tau_2} 
e^{- \frac{\tilde v}{2} (\underline{B}_1 - \underline{B}_2)^2 (\tau_0 - 
\tau_2)} ,
\end{equation}
and interchanging integrations in Eq.~(\ref{eq:45})
\begin{equation}
  \label{eq:50c}
 \int^{\tau_0}_0 \, d \tau_2 \,
 \int^{\tau_2}_0 \, d \tau_1 \, =
 \int^{\tau_0}_0 \, d \tau_1 \,
 \int^{\tau_0}_{\tau_1}\, d \tau_2 \, ,
\end{equation} 
a partial integration with respect to  $\tau_2$ is performed.
Together with the following equation (cf. Eq.~(\ref{eq:44})) 
\begin{equation}
  \label{eq:51}
\left(  \frac{\partial}{\partial \tau_2} +
 \frac{\tilde v}{2} \underline{B}_1^2 \right)
G(\underline{B}_1,\tau_2;\underline{B},\tau_1)\>=\> 
- i\tk\, \nabla^2_{\underline{B}_1} 
 G(\underline{B}_1,\tau_2;\underline{B},\tau_1),
\end{equation}
and a further integration by parts  in ${\underline{B}_1}$ similar
to Eq.~(\ref{eq:50a}),
the spectrum (\ref{eq:2.8}) is finally  expressed by
\begin{eqnarray}
  \label{eq:5x}
\frac{\omega d I}{d\omega dz d^2 \underline{U}} & = & 2
\frac{\alpha_s C_F}{ \pi^2 L}\,  {\tilde{\kappa}^2}
 Re  \, \int^{\tau_0}_0 \, d \tau_1 \, \,
 \int^{\tau_0}_{\tau_1} \, d \tau_2 \,
 \int \,{d^2 B_1}
e^{ i \underline{B}_1 \cdot \underline{U}} \, 
e^{- \frac{\tilde v}{2} \underline{B}_1^2 (\tau_0 - 
\tau_2)}
\, \nabla_{\underline{B}_1} \cdot \nabla_{\underline{B}}
 G(\underline{B}_1,\tau_2;\underline{B}=0 ,\tau_1) 
\nonumber \\
& + & \frac{\alpha_s C_F}{\pi^2 L}\, {\tilde \kappa}
 Re  \, \int^{\tau_0}_0 \, d \tau_1 \,
 \int \,{d^2 B_1}
e^{ i \underline{B}_1 \cdot \underline{U}} \,  
 \frac{\underline{U}}{\underline{U}^2} \cdot  \nabla_{\underline{B}}
G(\underline{B}_1,\tau_0;\underline{B} = 0,\tau_1) .
\end{eqnarray}
In the second term the transform
\begin{equation}
  \label{eq:tr}
 \int \,{d^2 B}_2 \,
e^{ -i \underline{B}_2 \cdot \underline{U}} \,
 \frac{\underline{B}_2}{\underline{B}_2^2}
 = - 2 \pi i \frac{\underline{U}}{\underline{U}^2}
\end{equation}
has been inserted.
 The subtraction terms at $\tk\!=\!0$  vanish. 

The two terms on the right hand side 
of (\ref{eq:5x}) correspond to and are identical
with the terms given by Eqs.~(A.5) and (A.6), respectively,
in the appendix of the paper by Wiedemann \cite{Wiedemann}.
These terms express the gluon radiation spectrum off a quark
produced in and propagating through the medium of size $L$.


\vspace{0.5cm} 
\subsection*{Acknowledgments} 
\noindent 
We thank Urs A.~Wiedemann for useful discussions and for
 pointing out the discrepancy between \cite{Baier} and \cite{Wiedemann}. 
The research  by R.~B. is supported in part 
by Deutsche Forschungsgemeinschaft (DFG), Contract Ka 1198/4-1.


\end{document}